\documentclass{aa}
\usepackage{graphics}

\begin{document}
\newcommand{\kms}{\,{\rm km}\,{\rm s}^{-1}}
\newcommand{\mum}{$\,\mu$m}
\newcommand{\muk}{$\,\mu$K}
\newcommand{\GHz}{$\,$GHz}
\newcommand{\mJy}{$\,$mJy}
\newcommand{\phm}{\phantom{-}}
\newcommand{\pho}{\phantom{1}}
\def\simgt{\mathrel{\raise0.35ex\hbox{$\scriptstyle >$}\kern-0.6em
\lower0.40ex\hbox{{$\scriptstyle \sim$}}}}
\def\simlt{\mathrel{\raise0.35ex\hbox{$\scriptstyle <$}\kern-0.6em
\lower0.40ex\hbox{{$\scriptstyle \sim$}}}}

\thesaurus{02(12.03.3; 13.09.1; 13.20.1; 11.05.2)}
\title{First detections of FIRBACK sources with SCUBA}
\author{Douglas Scott,$\!$\inst{1}\thanks{dscott@astro.ubc.ca}
Guilaine Lagache,$\!$\inst{2}
Colin Borys,$\!$\inst{1}
Scott C.~Chapman,$\!$\inst{1}\thanks{{\em Present address:}
Observatories of the Carnegie Institute of Washington, Pasadena,
CA 91101,~~U.S.A.}
Mark Halpern,$\!$\inst{1}
Anna Sajina,$\!$\inst{1}
Paolo Ciliegi,$\!$\inst{3}
David L.~Clements,$\!$\inst{4}
Herv{\'e} Dole,$\!$\inst{2}
Sebastian Oliver,$\!$\inst{5}
Jean-Loup Puget,$\!$\inst{2}
William T.~Reach\inst{6},
\and
Michael Rowan-Robinson\inst{5}}
\offprints{Douglas Scott}
\mail{Douglas Scott}
\institute{Department of Physics \& Astronomy, University of British
Columbia, Vancouver, B.C. V6T~1Z1,~~Canada
\and
Institut d`Astrophysique Spatiale, B{\^a}t. 121, Universit{\'e} Paris XI,
F-91405 Orsay Cedex,~~France
\and
Observatories of the Carnegie Institute of Washington, Pasadena,
CA 91101,~~U.S.A.
\and
Osservatorio Astronomico di Bologna, Via Ranzani 1, I-40127 Bologna,~~Italy
\and
Department of Physics \& Astronomy, Cardiff University, P.O. Box 913,
Cardiff CF22 3YB,~~U.K.
\and
Astrophysics Group, Imperial College London, Blackett Laboratory,
Prince Consort Road, London SW7 2BZ,~~U.K.
\and
Infrared Processing and Analysis Center, California Institute of Technology,
Mail Stop 100-22, Pasadena, CA 91125, ~~U.S.A.}
\date{Received date / Accepted date}
\titlerunning{SCUBA and FIRBACK}
\authorrunning{D.~Scott et al.}
\maketitle

\begin{abstract}
The FIRBACK (Far InfraRed BACKground) survey represents
the deepest extensive $170\,\mu$m
images obtained by the {\sl ISO} satellite.
The sources detected comprise about 10\% of the Cosmic IR Background (CIB) seen
by {\sl COBE}, and, importantly, were observed at a wavelength near the peak
of the
CIB.  Detailed follow-up of these sources should help pin down the redshifts,
Hubble types, and other properties of the galaxies which constitute this
background.
We have used the Submillimetre Common-User Bolometer Array (SCUBA) instrument
on the James Clerk Maxwell Telescope (JCMT) to search for sub-mm emission from
a sample of 10 galaxies which are bright at 170\mum\ and for which we had
accurate radio positions.  Statistically we detect this sample at both 450\mum\
(${\sim}\,4\sigma$) and 850\mum\ (${\sim}\,7\sigma$); individual objects are
convincingly detected in four cases at 850\mum, and one case at 450\mum.
Fits to simple spectral
energy distributions suggest a range of low to moderate redshifts, perhaps
$z\,{=}\,0$--1.5 for this FIRBACK sub-sample.
\keywords{Cosmology: observations --
Infrared: galaxies -- Submillimeter -- galaxies: evolution}
\end{abstract}

\section{Introduction}
In the far-IR/sub-mm waveband
there are currently two pressing (and related)
cosmological mysteries:
\begin{enumerate} 
\item what makes up the background detected by {\sl COBE}
(Puget et al.~1996, Fixsen et al.~1998, Hauser et al.~1998,
Lagache et al.~1999)?
\item how do sources detected in the sub-mm (see e.g.~Sanders~1999 and
references therein) relate to galaxies at other wavelengths?
\end{enumerate}
 
The study of
both issues provides important diagnostics of
galaxy formation models, and the
answers should help illuminate the dark ages of how the first objects
formed and subsequently evolved into present day galaxies.
The first question is the main motivation behind the FIRBACK project
(Clements et al.~1998, Lagache~1998,
Reach et al.~1998, Puget et al.~1999, Dole et al.~1999),
which made deep $170\,\mu$m images of the sky with {\sl ISO}, to resolve the
CIB into sources.  The second question has been the
main driving force behind the search for distant sources by several teams
using SCUBA, as well as the
large amount of follow-up work and comparison with other wavelengths
(e.g.~Smail, Ivison \& Blain~1997,
Barger et al.~1998, Blain, Ivison \& Smail~1998,
Holland et al.~1998, Hughes et al.~1998, Smail et al.~1998,
Barger et al.~1999, Blain et al.~1999a,
Chapman et al.~1999, Cowie et al.~1999, Eales et al.~1999, Lilly et al.~1999).
These observational campaigns are
untangling the problem of how obscuration by dust skews
our optical view of the early Universe, unveiling the `dark-side'
of galaxy formation out to distant redshifts, and helping provide unbiased
estimates of the global star formation history of the Universe.

In this letter we present the first results of a
study at the interface between these two puzzles.
Rather than trying to find new sources with SCUBA, we have carried out
450\mum\ and 850\mum\ photometry on sources already detected at
170\mum\ with ISOPHOT.

\section{The FIRBACK Survey}
The lifetime of the Infrared Space Observatory ({\sl ISO}; Kessler et al.~1996) 
is long over, and we now have in hand
the best information we will have from long-IR wavelengths until the launch of
{\sl SIRTF}.  For SCUBA observations, the smallest extrapolations come at the
longest wavelengths attainable with the ISOPHOT instrument (Lemke et al.~1996).
ISOPHOT is an imaging photo-polarimeter with 92 arcsec pixels and 1.6 arcmin
FWHM at $170\,\mu$m;
reasonably high signal-to-noise sources in `dithered' images
can be located with an accuracy of ${\sim}\,40$ arcsec.
Following the discovery of the CIB, {\sl ISO} was in a unique position to
investigate the sources which comprise this background.  Consequently
deep ISOPHOT images were obtained at
$170\,\mu$m of three separate ${\sim}\,1$ square degree regions of the sky,
selected for their low foreground emission -- the FIRBACK Survey.
One region, the `Marano' fields,
is only accessible from the southern hemisphere, while the other two
coincide with the `N1' and `N2' fields of the ELAIS project (Oliver~1997),
which used ISOCAM at $7\,\mu$m and $15\,\mu$m and ISOPHOT at 90\mum\ (although
only a fraction of the FIRBACK sources were detected at these other
wavelengths). These northern regions have also been mapped with the VLA
(Ciliegi et al.~1999).  In addition an area towards the
Lockman Hole is being studied by another group (Kawara et al.~1998).

Here we have concentrated on a sample of objects from the roughly two
square degree `N1' field, centred
at $16^{\rm h}\,11^{\rm m}$, $+54^\circ\,25^{\rm m}$.
For this first attempt at JCMT
follow-up we chose objects with secure, unconfused radio identifications,
relatively strong 170\mum\ emission, and additionally high 170\mum:21\,cm
flux density ratios (see Lagache et al., in preparation).
This last criterion was aimed at biassing the sample away from the lowest
redshift galaxies, and hence towards those which might have the highest
850\mum:170\mum\ flux density ratios.  We expect that with the addition of
more follow-up observations it should be possible to select future sub-samples
with a higher likelihood of being strong SCUBA emitters.

\section{JCMT Observations}
The data were obtained on the nights of 18--23 March 1999 using the
SCUBA instrument (Holland et al.~1999) on the JCMT.  The short- and
long-wavelength arrays were used simultaneously, at 450\mum\ and
850\mum, respectively.  We used `photometry' mode, chopping at the standard
7.8125\,Hz, and also nodding every second by
about 45 arcsec in coordinates fixed to the array (i.e.~there was no sky
rotation).  This means that each measurement
is a double-difference between the central bolometer and positions 45 arcsec
each side, corresponding approximately to positions of other bolometers
on both the long- and short-wavelength arrays.
The data were analyzed using the SURF package (Jenness \& Lightfoot~1998).
The raw data for the two arrays were
flat-fielded, corrected for extinction, had bad bolometers
removed, and had the average sky removed at each time interval.
The information from the off-beams was then added, assuming that one long
wavelength bolometer had an
efficiency of exactly 0.5, with the long wavelength bolometer on the other
side, as well as the two short wavelength off-beam bolometers, having slightly
lower values (see Borys et al.~1999, Chapman et al.~1999, for more details).
Adding the weighted off-beam signal always decreased the noise,
and generally increased the signal to noise ratio (SNR), but we carried out
the same procedure even when it slightly lowered the SNR.

Calibration was performed a few times per night using planets and other
strong sub-mm sources, and the values
we used were similar to the standard gains.
The standard deviation of the calibrations was 10\% at 850\mum\
and 12\% at 450\mum; these should be a reasonable estimate of the
uncertainty in the calibration.
At the low SNRs at which we are working,
the calibration uncertainty is not a major contributor
to the total uncertainty, and has essentially no effect on the SNR itself.

\begin{table}
\begin{tabular}{lccccr}
\noalign{\smallskip}
Source  &    $S_{170}$ & $S_{450}$ & $S_{850}$ & \\
        &    (mJy)     & (mJy) &  (mJy) & \\

N1-008  & 433    & $17.2\pm10.8$ &    $\phm3.3\pho\pm0.9\pho$ & $3.6\sigma$\\
N1-015  & 219    & $\pho2.8\pm18.3$ & $\phm2.6\pho\pm1.4\pho$ &  ${<}\,4.9$\\
N1-025B & 200    & $40.5\pm23.8$ &    $\phm2.4\pho\pm1.2\pho$ &  ${<}\,4.4$\\
N1-034  & 153    & $83.4\pm25.8$ &    $-1.0\pho\pm1.2\pho$    &  ${<}\,1.8$\\
N1-035  & 151    & $\pho8.6\pm21.0$ & $\phm0.8\pho\pm1.3\pho$ &  ${<}\,3.1$\\
N1-038  & 148    & $22.8\pm13.5$ &    $\phm6.0\pho\pm1.1\pho$ & $5.4\sigma$\\
N1-045  & 139    & $17.8\pm22.7$ &    $\phm2.1\pho\pm1.2\pho$ &  ${<}\,4.1$\\
N1-061  & 121    & $32.3\pm25.3$ &    $\phm4.5\pho\pm1.3\pho$ & $3.4\sigma$\\
N1-063  & 120    & $67.4\pm24.2$ &    $\phm4.8\pho\pm1.2\pho$ & $4.0\sigma$\\
N1-087  & \pho93 & $\pho8.3\pm20.8$ & $\phm1.2\pho\pm1.2\pho$ &  ${<}\,3.3$\\
\noalign{\smallskip}
{\bf Mean} & 178 & $23.8\pm\pho5.8$ & $\phm2.78\pm0.37$ &\\
\end{tabular}
\caption{FIRBACK identification (see Lagache et al., in preparation), together
with 170\mum\ flux density (with estimated systematic uncertainty of perhaps
30\%).
The 450\mum\ and 850\mum\ measurements are from our new SCUBA observations.
The final column lists the SNR for the ${>}\,3\sigma$
detections, and the 95\% Bayesian upper limits (in mJy) for the others.}
\label{tab:obs}
\end{table}

\section{Individual Objects}
At 850\mum\ we detected one source at ${>}\,5\sigma$ and a further three
at ${>}\,3\sigma$.  While we would not claim detection of the
other sources, they generally have positive flux density (see next section),
and certainly there are good upper limits in each case.  This last remark
applies to the 450\mum\ data,
where there is only one detections above $3\sigma$.

Bayesian 95\% upper limits can be obtained for all our non-detections,
by integrating a Gaussian probability, neglecting the
unphysical negative flux density region.  The 850\mum\ upper limits for our
six non-detections are given in the last column of Table~\ref{tab:obs}.
At 450\mum\ the limits are generally
less constraining for reasonable SEDs, being around $30\,$mJy in the best
cases.

Using a combination of 170\mum, 450\mum\ and 850\mum\ data, we can place
some constraints on the spectral energy distributions (SEDs) of these FIRBACK
galaxies.  Assuming a grey-body spectrum, we can obtain a limit on some
combination of luminosity, temperature, spectral index and redshift.
Here we choose to normalize the luminosity at 170\mum, and then use the
SCUBA data to constrain the redshift.  To do this we assume standard
values for the dust temperature, $T_{\rm d}\,{=}\,40\,$K
(typical for sub-mm selected galaxies, e.g.~Blain et al.~1999b,
Dunne et al.~1999), and spectral index
of the dust emissivity, $\beta\,{=}\,1.5$.
Because we do not know the absolute luminosity of any source,
our results are degenerate in the ratio $T_{\rm d}/(1+z)$, and so we
are unable to tell apart cooler objects at lower redshift from hotter
objects at higher redshift.

Assuming a uniform prior distribution of
redshifts, we can obtain a Bayesian 95\% confidence range on the redshift
implied for each source by our 450\mum\ and 850\mum\ data,
where it is understood that the redshifts
can be scaled using a different dust temperature such that
$(1+z)/T_{\rm d}$ is held constant.
We find that the objects with
the highest 850\mum\ flux densities have the highest implied redshifts:
e.g.\ $0.66\,{<}\,z\,{<}\,1.23$ for N1-038.  Those with lower flux densities
at 850\mum\ generally only yield upper limits to the redshift: e.g.\
$z\,{<}\,0.55$ for N1-015.  There are no 
objects for which we would infer $z\,{>}\,2$.  However,
higher redshifts could be accommodated by adopting a higher dust
temperature or a higher value of $\beta$ (and lower redshifts by lowering
these parameters).

The object in our sub-sample which is brightest at 170\mum, N1-008, is
hard to fit with any reasonable SED; it has approximately half the
850\mum\ and 450\mum\ flux density expected from even a $z\,{=}\,0$ source
with the same 170\mum\ flux density.  This suggests that there might be
more than one source contributing to the ISOPHOT flux, which is not
particularly unlikely, since the FIRBACK Survey is operating near the
confusion limit.  On the other hand, optical images show little in the
error circle except for a very bright obvious spiral galaxy, and so from
that point of view this is not a case where we expect more than one
source in the ISOPHOT beam.
Another possibility is that our JCMT beam did not include all the flux, since
optically this object appears quite extended.  However, we would expect the
sub-mm emission to be more concentrated than the optical, and hence we are
likely to have included most of the dust emission within the beam.
This object will be discussed more extensively in Lagache et al.~(in
preparation).

Relatively poor values of $\chi^2$ are also found for N1-034 and
N1-063.  These result either from low SCUBA flux densities relative to
ISOPHOT, or somewhat high SCUBA 450\mum\ flux densities compared with 850\mum.
Higher SNR data, or data at additional wavelengths are required to
determine  whether these flux densities come from multiple objects, complex
SEDs, or simply the low SNRs of the current data for these objects.

Another way of estimating the redshift relies on the radio/far-IR
correlation (e.g.~Helou, Soifer \& Rowan-Robinson~1987,
Carilli \& Yun~1999, Barger, Cowie \& Richards~1999,
Smail et al.~1999).
Using the 20\,cm VLA data from Ciliegi et al.~(1999) and the explicit
correlation (using $\beta\,{=}\,1.5$, $T_{\rm d}\,{=}\,40\,$K)
from Carilli \& Yun (1999),
we find $z\,{=}\,0.6$, 1.2, 1.1 and 1.4
for N1-008, N1-038, N1-061 and N1-063, respectively (our four
850\mum\ detections).  Except for N1-008 these are in broad
agreement with the values obtained from the sub-mm and far-IR data alone.

\begin{figure}
\resizebox{\hsize}{!}{\includegraphics{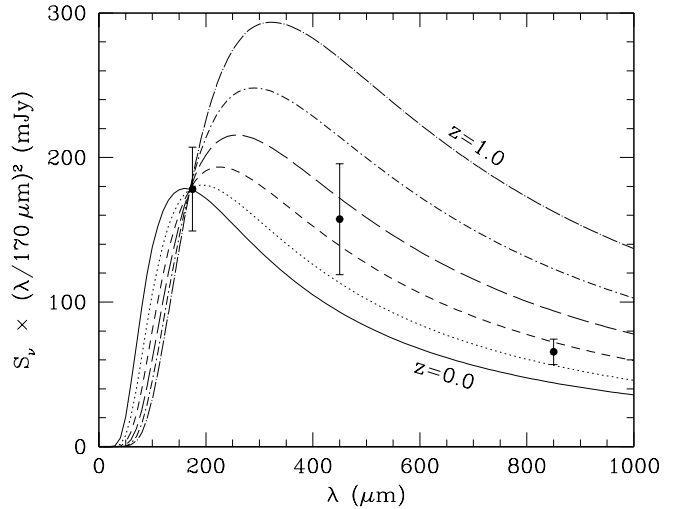}}
\caption{Average spectral energy distribution for our
FIRBACK sub-sample.  The 170\mum\ point is
the average of our sub-sample of 10 FIRBACK sources, with the error-bar
here being the standard error from the scatter among them.
The 450\mum\ and 850\mum\ points
are from our new SCUBA observations.  For the sake of visual clarity we
have multiplied the $y$-axis by $x^2$.  The curves show emission from
modified blackbodies, normalized to the 170\mum\ flux density, with
$T_{\rm d}\,{=}\,40\,$K and $\beta\,{=}\,1.5$, and for $z\,{=}\,0.0$
(solid line) up to $z\,{=}\,1.0$ in steps of 0.2.
Note that the shapes of these curves are degenerate in the combination
$(1+z)/T_{\rm d}$.}
\label{fig:sed}
\end{figure}

\section{Statistical Results}
If we combine all the data together, then we can obtain a much more
precise picture of the average galaxy in our FIRBACK sub-sample.
The average flux densities are given in Table~1, and represent a $4.1\sigma$
detection at 450\mum\ and a $7.3\sigma$ detection at 850\mum.  The average
SED is shown in Figure~1.  It is clear that for $T_{\rm d}\,{=}\,40\,$K
the best fitting redshift is around 0.3, with values as low as $z\,{=}\,0$
or higher than $z\,{\simeq}\,0.6$ providing relatively poor fits.
However, we can certainly
accommodate $z\,{\sim}\,1$ galaxies for higher temperatures, and $z\,{\sim}\,0$
galaxies for lower temperatures.  Since higher dust temperature is
seen in star-bursting galaxies, compared with more typical star-forming
galaxies, then it is possible that a fraction of these sources are
much more luminous and at higher redshift.  However, it would take
unrealistically high temperatures, $T_{\rm d}\,{\ga}\,100\,$K, to push some of
the objects up to say $z\,{\simeq}\,3$ (although the possibility of
gravitational lensing could complicate this).

Models of galaxy populations (e.g.~Guiderdoni et al.~1998),
designed to fit number counts at a range of wavelengths, as well as the CIB,
predict that the average FIRBACK galaxy is indeed at $z\,{\sim}\,1$.  The other
possibility is that we have uncovered a new population of relatively
nearby star-forming galaxies, which do not show up clearly in surveys
at other wavelengths.  The degeneracy
between $(1+z)$ and $1/T_{\rm d}$ makes it impossible to decide between these
possibilities using the
far-IR and sub-mm data alone.  Although it seems unlikely that all our
objects can be at low redshift, this could obviously be resolved by
obtaining optical redshifts.  Once redshifts have been obtained, then
the sub-mm data will be invaluable
in measuring the properties of the dust in the various galaxy types:
far-IR luminosities, dust temperatures and emissivities.

\section{Conclusions}
We have carried out the first SCUBA follow-up of FIRBACK sources.  We found
that they are generally detectable in the sub-mm; those with somewhat
higher 850\mum\ flux density may be at $z\,{\sim}\,1$, while those which are
fainter in the sub-mm may be more normal galaxies at $z\,{\sim}\,0$.
Models of evolving galaxy populations which provide a good fit to the
170\mum\ counts, as well as counts at other wavelengths
(e.g.~Guiderdoni et al.~1998) predict that the median redshift of the
FIRBACK galaxies is around 1 (Puget et al.~1999).
Our results are consistent with this, provided that the average galaxy in
our sub-sample is a distant star-bursting galaxy with fairly hot dust
temperature.  The other possibility
is that some could be from an otherwise unknown population of low redshift
star-forming galaxies with relatively low dust temperature.  Further
observations at sub-mm and other wavelengths should decide this issue,
and reveal the detailed properties of the galaxies which
comprise the CIB.

\begin{acknowledgements}
This work was supported by the Natural Sciences and Engineering Research
Council of Canada.
The James Clerk Maxwell Telescope is operated by
The Joint Astronomy Centre on behalf of the Particle Physics and
Astronomy Research Council of the United Kingdom, the Netherlands
Organisation for Scientific Research, and the National Research
Council of Canada.
\end{acknowledgements}

\end{document}